# Scaling agile on large enterprise level – systematic bundling and application of state of the art approaches for lasting agile transitions

Alexander Poth
Volkswagen AG
D-38440 Wolfsburg,
Germany
alexander.poth@volkswagen.de

Mario Kottke
Volkswagen AG
D-38440 Wolfsburg,
Germany
mario.kottke@volkswagen.de

Andreas Riel
Grenoble Alps University
G-SCOP Laboratory
F-38031 Grenoble,
France
andreas.riel@grenoble-inp.fr

*Abstract* — **Organizations are looking for ways of establishing agile and lean process for delivery. Many approaches exist in the form of frameworks, methods and tools to setup an individual composition for a best fit. The challenge is that large organizations are heterogeneous and diverse, and hence there is no "one size fits all" approach. To facilitate a systematic implementation of agile and lean, this article proposes a transition kit based on abstraction. This kit scouts and bundles state of the art methods and tools from the agile and lean community to align them with governance and compliance aspects of the specific enterprise. Coaching of the application of the transition kit ensures an adequate instantiation. The instantiation handles business domain specific aspects and standards. A coaching governance ensures continuous improvement. An example of the systematic application of the transition approach as well as its scaling is demonstrated through its application in the Volkswagen Group IT.**

## I. INTRODUCTION

THE diversity of an enterprise's business areas demands individualized implementations of lean and agile. Often the main goal of the agile transition is to gain delivery speed. According to Albert Einstein: "Make everything as simple as possible, but not simpler", we have to find a way to achieve effectively the simple yet complete organizational setting. Furthermore, Conway's law [44] leads us to develop something customizable to build a lean and agile organization for a best fit to the specific products and services, which the organizational unit creates and delivers. These two aspects have to be handled to realize a lasting and sustainable transformation.

Large established enterprises are built around different business areas with independent business units or divisions in a matrix structure [1]. Most of these business units have the size of a medium-sized enterprise. Furthermore, large enterprises are mostly based on large delivery pipelines oriented on the efficiency paradigm of the Taylorism [45]. Any transition aid for application within such context has to be able to handle this setting. More specifically during our first operational coaching of projects within the Volkswagen

☐ This work was not supported by any public organization or funding

Group IT in past transformation initiatives we identified the following aspects an agile transition aid has to address:
1) Identify the target organization for the transition, including its boundaries.
2) Identify the organization's value stream, including interfaces at the boundary to "external" partners.
3) Define and clarify the transition's objectives.
4) Evaluate different approaches to lean and agile for their suitability in the particular organizational context.
5) Implement the selected approaches:
   - Train people in the approach.
   - Re-organize the workflows according to the approach.
   - Align the new setting with the enterprise's governance and compliance structures.
6) Install cyclic checks for transparency and improvement:
   - on a local view of the transition for "self-optimization";
   - on a global enterprise view to develop the "setting";
   - offer an open networking platform to reflect transitions.
7) Support scaling of transitions

This leads to the investigation question: How is it possible to address these demands with an easy to handle approach, which can be applied by a team of coaches in a structured fashion? Our objectives for achieving this are the following:

(O1) A transition kit is needed that is able to handle lean and agile approaches.

(O2) Based on the organization's stakeholders' current mindsets a specific set of methods and tools for the workflows has to be implemented.

(O3) The organization's specific product setting has to be taken into account appropriately.

## II. RELATED WORK

This section investigates related published work with a focus on a holistic approach to addressing those. There is a huge amount of relevant approaches to organizational development [2], alternative setups like holacracy [3] or transitions [34] starting on grounded theories [32] to practice collections of other enterprises [33]. We are interested in identifying well-known approaches, methods and tools that can be used as a kind of reference in various settings to reduce complexity. Our contribution is to bring together the team setting with its cultural and mental history thanks to an adequate set of approaches, methods and tools to realize a effective and sustainable transition. We structured related work according to this scope, rather than elaborating on all kinds of available methods and tools at the time of writing this article.

### A. Setting Analysis

The Cynefin [5] and the Stacey-matrix [7] are approaches to classify the product context into a complexity setting and the drivers of the transformation [36]. This are useful approaches to identify the development context of the transitions product environment. The spiral dynamics model [4] and the Group Development Questionnaire (GDQ) [8] classifies the maturity of a group of humans who focusing together on an objective or purpose. As setup point on the teams maturity for transition approaches and methods this is crucial. Value-stream mapping [6] is an approach to optimize processes in a given setting especially for software [35] which come for the production [46] to the software development [47].

### B. Lean and Agile Approaches

Scrum [13] and XP [15] are team approaches focusing on agile working. Kanban [14] works in a team and in bigger organizations. SAFe [9], LeSS [10], Nexus [11] and Scrum@Scale [12] are approaches to handle the synchronization of more teams in a bigger organization. Furthermore a lot of variants are existing like Disciplined agile delivery (DAD [48] or Agile modeling (AM) [49].

### C. Methods and Tools

Design Thinking (DT) [16] is a method to develop an initial product in an iterative hypothesis based manner. Minimum Viable Product (MVP) [17] and derivations like Simple Lovable and Complete (SLC) [18] are tools to define an initial product version for delivery. Business Model Canvas (BMC) [19] or Lean Canvas [50] and its variants like for organizations internal communication [20] are used to identify the setting of a business to optimize in a later step the value-stream for product and its revenues. The Product Vison Board (PVB) [21] is used to for focusing a team on a product. INVEST [22] is used to systematically identify requirements for a product. Definition of Done (DoD) [23] or derivations like Levels of Done (LoD) are used to ensure that product versions fit quality definitions. To keep the delivery procedure lean and focused Product Quality Risk (PQR) [24] mitigation can guide to the delivery.

### D. Organizational culture and team psychology

The culture moves to a more internal lean start up [26] setting also in bigger enterprises. The objective of most digital business models [19] is scaling into the mass-markets [25]. Coaching approaches are reflected to be effective in the setting [27] to address the agile teams.

### E. Governance, Risk and Compliance

Governance has to establish standards like ISO 9000 for quality management, standards for risk management like the ISO 31000 and additional domain specific standards. Approaches for agile risk handling exists [31]. For service management, the ISO 20000 is an established anchor. Some concepts for agile governance [28] and [29] exist, however their scope is limited to applying agile or lean principles outside a globally acting [30] enterprise context.

## III. TRANSITION PROCESS

Within Volkswagen Group IT, we do not use one given method, model or tool because the organizations' s size demands context adequate approaches. More than 2000 internal employees and a lot of divisions and organizational units indicates the complexity which the transformation has to deal with. Therefore we decided to start with the basis: the team.

A transition kit and process has been developed and maintained by a central team, the Agile Center of Excellence (ACE), which guides and coaches agile transitions. ACE is a department within the Group IT uniting initial agile users from the first agile projects. The transition process consists of three phases: the transition itself, as well as a pre- and post-transition phase to ensure sustainable transitions. ACE supports transitions in the Group IT and other business areas of the Volkswagen AG based on their transition process and kit that has been enhanced over years.

The coaches establish the initial setup and alignment during the coaching phase of the team's external process expectations (figure 1). This is the initial link to process safety and compliance for the teams. The long-term alignment is checked by the project review.

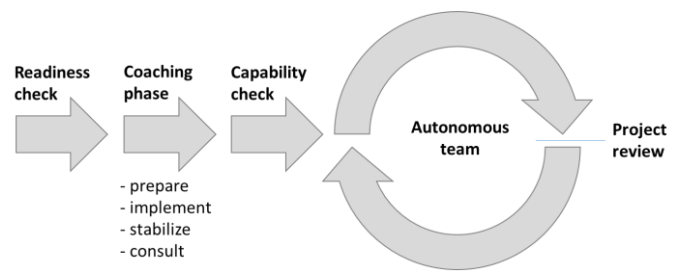

Fig. 1: Coaching to team autonomy with integrated compliance check

In the pre-transition phase, the "readiness check" is conducted to identify the status quo and objectives of the transition. The status quo identifies roles like sponsor of the transition, product/business owner and the team setting. Furthermore, agile artifacts like for the backlog and its items are investigated. Based on the evaluation of the acquired information, a transition can be recommend or not. In case of recommendation, the ACE can support the transition with the transition kit. In case of a non-recommendation to start a transition, the ACE will not support because there is low chance to finish the transition in time successfully. The biggest challenge during this phase is the interlocutor's honesty. All transition aspects are based on it and conveying information honestly and completely is needed to give the transition a chance to be successful. Therefore we decided that we start the transition with motivated and voluntary units supporting the transition and meeting the prerequisites from the outset.

The main purpose of the transition kit is to enable teams to deliver most product benefit within in a continuously changing environment. The ACE coaches help start agile projects and teach the team how to deal with impediments. Additional ACE tasks are:
- first aid in network,
- promoting agile methods,
- connecting committees,
- supporting knowledge transfer,
- combining agility practices of brands,
- enable leadership to act in an agile way,
- sensitize the unit to get an agile mindset,
- pay attention to process safety.

Every transition phase starts with a contract clarification to get a clear understanding of what will happen. Referring to the Agile Manifesto [39], the contract does not describe the HOW, but rather the WHAT. Depending on the results of the "readiness check" and the needs of a team, product or project, the transition duration will be estimated and a coaching package will be offered (cf. Section V). The contract defines the purpose, deliverables from both sides and the organizational issues like contractor and cost issues. The transition itself has four steps:
1. Preparation (evaluation of team and product setting)
2. Implement the methods and the tooling
3. Stabilization
4. Consulting

The *preparation* includes the execution of a kickoff workshop, consulting (project leads, development team) and agile workflow creation. Also includes support, moderation, preparation of the management and creation of Definition of Ready/Definition of Done and initial product backlog with the team. The initiation of the first meetings like refinement, planning, review and retrospective is a task, too.

To *implement the methods and the tooling* the guide is always available for the team. The coaches train the team and the roles inside e.g. Scrum Master, Product Owner etc. to do the job to be done. The guide also moderates the necessary meetings like review, daily, retrospective, planning or refinement. Furthermore the guide assists the change management for motivation, conflict solving and workflow changes. The coaches are instantiating the initial setup and alignment of team external process expectations. This is the initial link to process safety and compliance for the teams. The long-term alignment is checked by the project review of the post-transition phase.

The *stabilization* step during the coaching (figure 1) is not so intensive for the coaches because the team should do their first steps alone. The coaches are always available for support and assistance, and in special cases will also assume the role of moderators. In this step, their job is to motivate, inspect, adapt and strengthen the change to be sustained. Solving conflicts is also part of it.

The *consulting* step is demand driven and mostly the end of the transition phase (figure 1). If the customer needs help, the coaches will help and give answers for questions to events, roles and workflow. The guides help the change management manage conflicts and adapt innovations.

The post-transition phase starts with a hold back (capability check in figure 1) of the transition team during the stabilization step and ends with a report. The report reflects the coaching contract objectives and also the agile issues and elements. Furthermore, the team or organization is registered as "agile". This flag will be used for the future agile governance checks (cf. Section VI) to ensure sustainability of the transition and incremental development of the people to stay up to date about the state of the art about agile.

## IV. TRANSITION KIT

For the demand of the Volkswagen Group IT to transform classic project management to business agility we developed the transition kit. It contains the methods and tools which are released during the transition process. Within the transition process, we try to find the best choice of approaches, methods and tools to create value faster. The transition kit addresses the implement step of figure 1. The transition kit focusses on the key parts of figure 2. These key parts are the product or service which is the delivery to the customer, the team realizing and supporting the products, as well as the governance ensuring organizational standards. Governance can also be triggered by external demands for example from legislative changes. The transition kit has to support the setup of the demanded skills and capabilities of the team from the outcome view (product/service). Furthermore the governance has to handle the product or service risks by guiding the teams to be able to balance the business value and risks related to the product or services they handle.

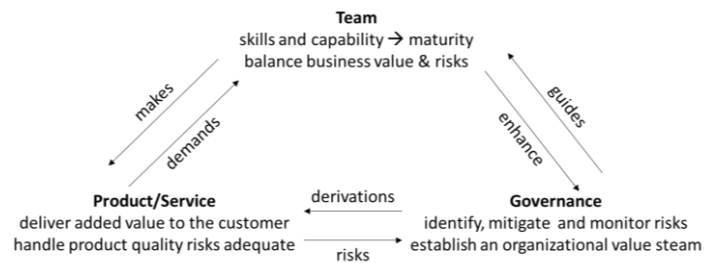

Fig. 2: Transition's key parts and their relationships

All three parts interact and need a holistic handling by the transition kit to realize a comprehensive product or service from the customer view who is using the product/service. The tool selection of the transition kit (table 1) is initially based on a first fit for purpose. This first fit was realized by a literature review [40] to identify artifacts for the initial transition kit. The transition kit contains approaches, methods and tools which helps the coach and team to go in an effective way into the right direction during the transition. Over the life cycle the transition kit will be enhanced by adding and changing artifacts to better fit the current organizational culture, for an easier integration into the coaching or simpler use in a self-service approach for teams without coaches. The enhancement is triggered by feedbacks. While everybody can suggest new artifacts for the transition kit, the ACE will evaluate and integrate relevant suggestions during their cyclic inspections. The objective is not to have a maximum of possible elements in the transition kit, but rather to have a lean transition kit that can be trained easily and is effective in most organizational settings. To make it easy to find the right artifacts the transition kit is aligned with the product complexity, team maturity and the agile approaches.

To identify the projects the ACE supports with coaches we use the Stacey matrix. It is an easy to use way to identify if agile is helpful or not.

The assignment of tools to phases is based on experience during the supported transitions. The determination of the appropriate transition kit artifacts is done according to the following procedure: To start in a value-driven way, the initial focus of the transition is the product or service. The product is located on the Stacey-matrix. Over the product life-cycle, the complexity location is more or less stable in emerging markets – with a trend to reduction of complexity in mature markets or at the end of a product life-cycle. The current state is identified and the future result or objective will be considered to advance in the right direction. In a second step, the relevant governance guidelines are identified. Based on the product and governance demands, the current team skills and capabilities are focused on. The product team setting is located in the spiral dynamics model (table 2) color levels. This location is important because often organizations coined by Taylorism established over years, act on the "red level". These teams have to make their mindset leaner to achieve the "blue level". Agile teams typically act on levels of blue and higher. Each team has to grow level by level in their maturity. This leads to the adaptation of the used artifacts over the maturity journey of a team. Based on the team's maturity and their product environment complexity, the appropriate agile approach will be selected mostly based on the suggestions of table 2, however the guide and the team can make adjustments if they think another artifact would fit better. The artifacts help the team to progress in the transition, but most of the transition effort is to enable and coach the team to deliver a product. Some examples about the experience-based labeling of the table: Why is Kanban applicable in beige teams? Kanban does not define a set on rituals like retrospectives from Scrum which demands a minimum level of trust in the team to discuss issue frankly. Kanban itself is a more "mechanical" approach. Both approaches can be used to develop the teams to higher levels. With higher levels the teams are acting different within the same approach by discovering more opportunities with the higher team trust and openness. Why do we have small "item" like MVP and "big items" like Safe in the table? Depending on the context it is useful to start with small items to support individual transitions of teams. In case of a more top-down demand a big item reduces discussions about how to start because it is like a pre-defined "package" ready for rollout. This is also the reason why the transition kit does not add every approach, method or tool – it selects some (first fit algorithm based) which work in the industrial context and tries to reduce redundancy were it is useful and possible by offering enough variance for the individual coaching of teams. The objective for the transition kit is to offer a practicable way for the transition of a team, without proposing any way possible.

TABLE I.
TRANSITION KIT ARTIFACTS AND THEIR MAPPING TO TRANSITION SPECIFIC KEY-ASPECTS

| Method/tool | Spiral dynamics team maturity | Stacey | Phase (average) | Application |
|---|---|---|---|---|
| Retrospective | Purple or higher | All | pre, mid, post | High (over 75%) |
| Design Thinking | Blue or higher | All | Pre | Low (under 25%) |
| Minimum Viable Product (MVP) | Orange or higher | Complex & complicated | Pre | Mid (25% to 75%) |
| Simple Lovable and Complete (SLC) | Blue or higher | Complex & complicated | Pre, mid | Low |
| Business Model Canvas (BMC) | Purple or higher | Complex & complicated | Pre | Low |
| Product Vison Board (PVB) | Purple or higher | Complex & complicated | Pre | Low |
| INVEST | Purple or higher | Complex & complicated | Mid | Mid |
| Definition of Ready (DoR) | Blue or higher | All | Pre, mid | Mid |
| Definition of Done (DoD) | Blue or higher | All | Pre, mid | Mid |
| Levels of Done (LoD) | Blue or higher | Complex & complicated | Mid | High |
| Product Quality Risk (PQR) | Ref or higher | Complex & complicated | Mid | Low |
| Scrum | Purple or higher | Complex & complicated | Pre, mid | Mid |
| Extreme Programming | Green or higher | Complex & complicated | Pre, mid | High |
| KANBAN | Beige or higher | Complex & complicated | Pre | Low |
| SAFe | Red or higher | Complex & complicated | Pre, mid | Mid |
| LeSS | Blue or higher | Complex & complicated | Pre, mid | Low |
| Nexus | Orange or higher | Complex & complicated | Pre, mid | Low |
| Scrum@Scale | Orange or higher | Complex & complicated | Pre, mid | Low |

TABLE II.
MATURITY LEVELS OF THE SPIRAL DYNAMICS MODEL [4]

| Name | Structure | Motives | Characteristics |
|---|---|---|---|
| Beige | Loose bands | Survival | Archaic, instinctive, basic, automatic |
| Purple | Tribes | Magic, Safety | Animistic, Tribalistic, Magical, Mystical |
| Red | Empires | Power, Dominance | Egocentric, Explorative, Impulsive, Rebellious |
| Blue | Pyramidal | Order, right & wrong | Absolutistic, Obedient, Purposeful, Authoritarian |
| Orange | Delegative | Autonomy, achievement | Materalistic, Strategic, Ambitious, Individualistic |
| Green | Egalitarian | Approval, Equality, Community | Relativistic, Personalistic, Sensitive, Pluralistic |
| Yellow | Interactive | Adaptability, Integration | Systemic, Conceptual, Ecological, Flexible |
| Tortoise | Global | Compassion, Harmony | Holistic, Global |
| Orange or higher | Complex & complicated | Pre, mid | Low |
| Orange or higher | Complex & complicated | Pre, mid | Low |

The transition kit does not focus on finance procedures of the enterprise however some programs are using for example MVP based finance planning to manage their annual budgets in an agile fashion. However the approaches, methods and tools can be applied to special functions. For example, the Group IT security organization was an early adapter.

The transition kit is designed to develop culture, team maturity and products/services together. Of course it is possible to enforce some methods or tools on lower leveled teams, but the real opportunities are only realized within the right culture and team context. The application column in table 2 shows a current distribution of the application the line in teams.

V. COACHING

ACE offers different volumes of coaching packages [37]. The package size is defined by the amount of time a team gets support from the transition team. The intensity depends on the time the guides (coaches) support the team. The coach sets up the team to address the demands and objectives of the transition by using the transition kit as guidance framework for the transition. The main focus of coaching is on the events, mindset, team performance, roles and their tasks, the used methods and how to inspect and adapt. Therefore the guide will use workshops with the whole team, as well as direct coaching.

Every coaching starts with a collection of information. This is necessary to find out what the transition (e.g. the project or team) really needs. To implement agility, the coach starts creating awareness of agile principles and values. With growing understanding, the flow will be created to support agile behavior. This means that the team can welcome and handle requirement changes having influence on the actors. The coach helps to give the team the power and knowledge they need. This is an ongoing process during all transition phases and may not be finished when the coach leaves the team.

When the transition goal is clear, the coach has to decide on which level to be most effective. If the transition has most effect on teams, the coach will focus on team members. The objective of the coach is to start small and establish the simplest possible set of artifacts from the transition kit to realize the objectives of the transition. For instance, if the coach decides implementing Scrum, he will support the Scrum team including the Scrum Master, the Product Owner and the development team. If the transition requires an organizational change, the coach will spend more time on management level where the responsibility for the portfolio is located. The tools and methods are all based on values and principles. The coach's main task is to make clear what the effects of their actual application are. Furthermore, the coach facilitates the teams with methods and tools for generic product and service development. An example is requirements elicitation and engineering with the product vision board to align the requirements at least with epics and stories oriented with INVEST and PQR (cf. table 1).

VI. GOVERNANCE

Each enterprise needs a governance structure ensuring that fundamental things are done in a deterministic way, and at minimum according to the state of the art. The state of the art is defined by organizational settings or derived from the market standard and regulations. Consequently, also all agile and lean teams have to establish and ensure the state of the art for their products and services. Depending on the product specific aspects, on top of the state of the art additional factors have to be ensured, e.g. market advantages. During the coaching phase aligned with the transition kit this is delivered by a team external coach. The coach has to make the teams sensitive for this governance topic and their team responsibility to stay aligned in the future. After the coaching phase the teams are independent and have to care about the "update" to the developing state of the art on their own. To make it easier for the teams, the governance offers update information about state of the art changes, which can be adopted by the teams. However, the governance has to ensure the alignment with the rail guards and update them to fit the state of the art. Rail guards are typically artifacts ensuring that some basics are done by the teams like for example an approval evidence for a deployment. Furthermore, the governance has to verify the effectiveness of its settings. These effectiveness checks are realized with controls. Different (domain) standards for System and Organization Controls (SOC) like [42] exist, but all have in common that the effectiveness of the established procedures has to be adequately checked, and if needed an alignment action has to be triggered. To ensure alignment with the settings and the agile and lean mindset a project review is established [38]. The project review (see figure 1) checks different aspects of an agile team or organization. Depending on the project or product classification (based on risk etc.) it will be checked in a deterministic way or randomized picked for a review. This ensures a basic transparency of alignment with the state of the art of the current portfolio.

The reviews are conducted by some coaches who have been trained in the evaluation aspects and their rating criteria. This common understanding about the aspects and rating ensures comparable results to derive organizational issues. Furthermore, an objective is not to change existing review aspects to keep the historical results in the data-analysis pool.

The defined rail guards for the expected artifacts and outcomes for fulfilling external requirements like aspects of the GDPR [43] or quality standards like ISO 9000 are checked in the project review. The results are used on both levels, for the reviewed team as well as the overall organization. Most of the findings have to be addressed by the product teams, however some findings are seen in many teams. This is made by cyclic analysis of the project review results to identify "derivation pattern" which have to be addressed on the organizational level. A derivation pattern is identified if in a significant amount of the cyclic checks similar derivations are observed. This is the trigger to handle it not only on the specific product or service instance and start caring about it on a generic or organizational level. For each identified derivation pattern the governance checks why it does not fit to the product teams and their deliveries. This can lead to actions on the organizational level having a high bandwidth. Finally, there is the educational aspect that leads to inadequate setting – this is addressed by training or coaching offers to establish the things as intended. This may lead to refactoring the rail guards or artifacts to fit better into the project teams and the organizational culture. Figure 2 shows the relation between the product, the team and the governance. The relation "enhance" in figure 2 leads to the learning that as much as possible should be structured as self-service for the teams to reach higher autonomy and better scaling. This initial higher effort to develop the governance outcomes as self-service capability empowers the teams to live their self-organization and responsibility. To give feedback to the teams in a gamification context, the top ranked project review results are posted on an intranet page as a "champions league table" involving the entire organization.

The development and update of the transition kit is an additional important task to assure alignment with current regulations and the developing state of the art over the time. The transition kit has to support the governance artifacts like the rail guards during the team settlement. To do this, external and internal triggers are established. For example, the PQR method from the transition kit directly helps to make transparent why things are done in this way for some governance measures. An objective of the improvement of the transition kit from the governance perspective is to integrate as many measures as possible into the product or service artifacts or their direct production procedures. This integration makes it leaner and easier for the product teams to align their work with the expected outcomes and measures.

The Volkswagen Agile Community (AC) is the chance for everybody to get updates and the information about current development of agile and lean. It is an open community for networking and share knowledge about agile and lean. This includes also topics about the transition kit and agile governance.

DACH30 [41] is a trans-enterprise network to share experience about agile and lean. Trainings and skills are developed together. This ensures that the transition kit is reflected by external experts and is updated to the current insights of other enterprises.

The objective of the governance is to give the teams as much freedom for agility as possible while still demanding sufficient discipline from the teams to fit the compliance framework.

## VII. EXPERIENCE REPORT

At Volkswagen AG Group IT, the transition kit development started in 2016 to support the coaches' daily work and has been enhanced continuously by the ACE and the coach guild to address the challenges of migrating to lean and agile methods in a structured way. Currently more than 100 product/service teams and organizational entities have been coached based on the elements of the transition kit. All those elements have been deployed – some more often than others (see table 1, column application). The teams are from the Group IT as well as other areas of the Volkswagen AG like plant production planning or vehicle development organizations, as well as smaller organizations like board member offices. The teams are supported during the transition in different life-cycle phases of their products and service. Some teams started on a green field, some were already established delivery teams. The range of software developed by the coached teams covers a wide range – from standardized ERP systems supporting human resources and production logistics to special software for supporting specific intellectual property of a business area. Also the architecture differs from established 3-tier architectures to cloud native micro-service based systems. The coaching phase differs in time from a few weeks to many months – depending on the size of the team or organization. Additionally, within the Volkswagen AG there exist a number of self-service based transitions which are often unknown to the ACE. By using the self-service, the teams have a low entry barrier because they can do it on their own way and speed, but the risk of applying inadequate elements of the transition kit is higher without an experienced coach.

The following parts of the case study reflect the objectives O1 to O3 and the observations of the application of the transition kit in the coaching phase as well as the results of the project reviews to have a long term perspective on the sustainability of the transition.

The lean and agile approaches are mapped to the transition kit artifacts to support the artifact selection. Depending on the approach, more or less options are offered to be chosen by the coaches and teams (O1). There is a trend in smaller teams without an end to end responsibility to use Kanban. This is motivated by the external process dependencies which limit the team's autonomy and freedom. The teams are often part of process driven value chains which drive the cycle time and delivery-dependencies. Hence, sprint commitments are not easy for the team. On the other side there is a trend to SAFe for transitions of multi-team organizations. Both show that the upper maturity levels are often not achieved.

The maturity derived from the spiral dynamics model of the teams is mapped to the transition kit artifacts to support especially lower leveled teams by choosing adequate approaches. With higher maturity levels the transition kit gets less importance because the teams have the capability of improving on their own and develop their appropriate way with supporting methods and tools to address their specific situation best (O2). Many teams have started their transition from the red or blue level Taylorism driven culture. However, some teams are built from scratch and in a greenfield area. Here, a quick move to "higher" levels is possible, because they do not have to learn to forget established habits and culture. The coaches typically can see some progress of one or two levels during their supporting phase. In the project reviews after a longer time a further progress can be observed. But in case of no strict application of agile methods and mindset some teams also go down to their "roots" with Taylorism habits. For these teams a "refreshing" coaching phase is suggested, if they still want to become agile.

The specific product setting with the complexity and value stream is supported by the transition kit, too. The artifacts are mostly generic and fit to the typical product settings in the complex setting (O3). In the future it could be possible to simplify the transition kit more by substituting complexity specific artifacts by generic ones.

The fact that the agile teams investigated in the case study are not permanently co-located does not significantly impact the application of the transition kit because most of the teams have some cyclic common physical meetings like refinements or retrospectives and use in-between communication tools to setup virtual team rooms.

The case study identifies that all phases of the transition are applied and supported as intended by the transition kit as described in section IV. The transition kit makes it easier to for new coaches to deliver transition support in a project-style to the teams in a standardized way. The integration of the transition kit in the holistic enterprise environment with a centralized product delivery process compliance helps the coaches and teams to be effective also from a compliance perspective. The controls of the effectiveness work because some transitions were not started because the environment did not fit according to the results of the readiness check.

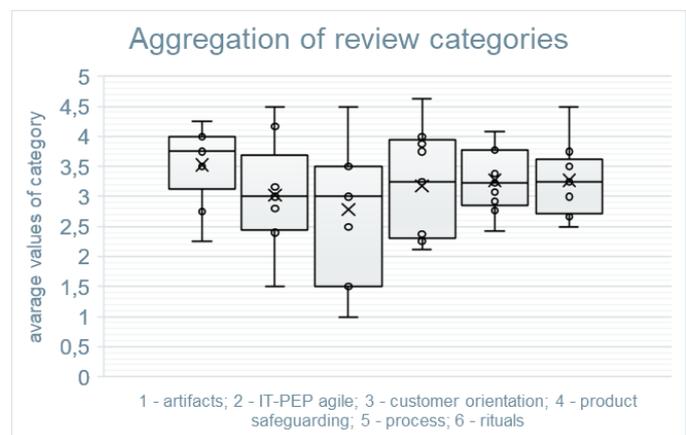

Fig. 3: Anonymized review results of the categories shows spreads and potentials (1 is most left bar – 6 most right bar)

The control project review with its check aspects helps to show the effectiveness of the transition and its sustainability in the teams later on (see Fig. 3). Based on these measurements and metrics for agile projects, agile processes, and agile teams the governance identifies improvement potentials. For example, one related to the agile process (IT-PEP agile which is the $2^{nd}$ bar in figure 3) effectiveness controls the re-thinking of the Group IT development process for a better alignment with agile and lean approaches and setting of guide lines which can easier integrated into operational excellence by the teams was indicated.

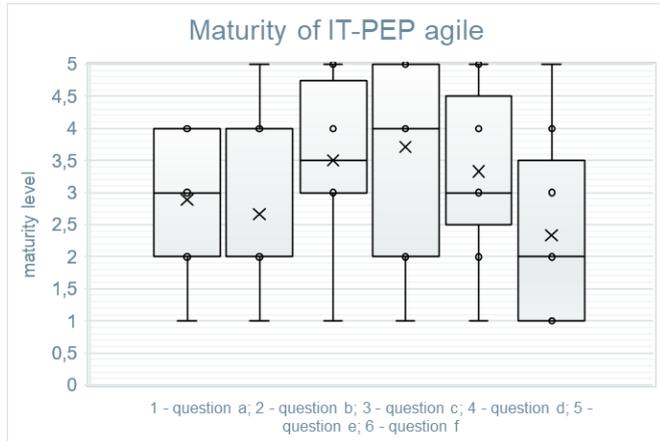

Fig. 4: Maturity of IT-PEP agile (1 is most left bar – 6 most right bar)

Figure 4 shows the results of category IT-PEP agile of representative project reviews between 2017 and 2019. The x-axis are checked aspects of the project review which is aligned the teams agile adaption and the governance aspects. A more detailed description of the aspects and their grouping on the x-axis is in [38] described. The y-axis shows the fulfillment of the checked aspect. The bar in the middle shows the $2^{nd}$ and $3^{rd}$ quartile of values. The trend on derivations to the standardized templates of the development process is visible (every question has low values and almost all also high values – especially question f in figure 4). This derivation has led to the creation of a community of practice as a kind of working group whose mission is to enhance the Group IT development process to be better aligned with the state of the art habits of agile and lean working teams. This is one way of feedback to improve the environment to be more agile.

Often the coaches also identify new approaches, methods or tools which are evaluated as a kind of experiment during a selected team coaching. Results and lessons learned from this experiments are reflected in ACE to improve the transition kit. Furthermore, the case study shows that some transitions are not lasting or sustainable. The effectiveness of the transition is checked by the review with a delay to the coaching phase. By comparing the results achieved during the transition with the results of the progress reviews the progress or back-steps of the teams can be made transparent and thereby used for deriving the appropriate improvement actions. The selection of the reviews was made from feedback applications by randomized picking from the successful team transformation list and high-risk labeled projects/products. The highest frequency is one year for conducting reviews in a team. This is to avoid too many reviews in short time periods by random picking without the chance for the teams to improve in between reviews.

VIII. CONCLUSION

The presented holistic scaling approach demonstrates that a centralized agile governance can help large enterprises scale agile transitions in the product and service teams. This centralized ACEs coach guild and Agile Community are used to manage the agile knowledge and enhance the transition kit. The setup of a self-service driven team governance is a chance for establishing a lean governance approach. Furthermore the lean and agile mindset in governance offers the teams the chance to participate in the future "look and feel" of the governance, such as the development of higher automation of governance tasks and their evidences. This automation objective is a logical consequence of the automation with the everything as code approach [51] of devops. The governance will check the effectiveness of the participation driven development with the controls like the governance initiated reviews to ensure that the enterprise enhance in a positive way aligned with the strategy. A second observation is that the governance develops fast if they live the lean and agile mindset themselves. Their responsibility is to serve the teams in an effective way to be compliant with external and internal requirements.

The evaluation about the effectiveness of coaching with a transition kit is seen on two points:
- At the end of the coaching phase on which the readiness check situation and the current outcomes of the capability check are compared.
- At the project review with the distance view (at least 1 year) after the transition coaching.

The objective of the ACE is to be effective by the coaching support. This is realized with the transition kit by applying and enhancing the transition kit continuously with the lessons learned from the transitions coaching. The efficiency is seen on the higher team transformation throughput of coaches. The issue is to have a generalized kit which is easy to instantiate in the specific team setting. This trade-off is a current enhancement focus of the transition kit. Furthermore a contribution is that this transition kit explicitly handles the mental team setting by application of the spiral dynamics model to apply adequate approaches and methods during the transformation to support effectivity the progress and sustainability also after the coaching phase.

IX. NEXT STEPS AND FUTURE WORK

Sustainability is a topic that needs more focus. Often the agile project review makes transparent that after the coached transition phase, the teams lose some of the leaned rituals etc. and fall back to pre-transition habits. We need to define or develop external triggers to reflect the team's rituals and progress in the developing of the agile and lean mindset without the coaches. This is a topic for an effective governance of the agile and lean processes.

Furthermore, the amount of skilled coaches does not scale with the demand. We need to enhance the transition kit to a complete self-service approach. Then teams with some "basic" skills can work more autonomously, needing less coaching. This is a governance and training issue. The training aspect is to enable the teams to do mostly everything in a self-service manner by offering a suitable transition kit. But on the other side the governance has to ensure that also self-service transitions have high quality outcomes.

Another open point is that the presented approach is only applied in a European enterprise culture. Its effectiveness in other cultural contexts still has to be investigated.

Next steps are the refactoring of the current process governance rail guards for a higher automation degree. The objective of the potential automation offers mature teams the integration into their automated product delivery pipeline (CI/CD chain). Some teams are currently experimenting and evaluating automated governance controls. The challenge is to find a balance between integrated standard tools and the freedom of the agile teams. Is automation an adequate indicator to determinate the product team maturity, especially in team's customized CI/CD chains? Will an individualized CI/CD chain slow down the integration of currently "independent" agile teams in future release trains of SAFe? Another interesting point is to extend the product based focus of the transition kit with a more lean and agile product finance scope like Beyond Budgeting [52].